\documentclass[%
reprint,
 amsmath,amssymb,
 aps,
]{revtex4-2} 
\usepackage{graphicx}
\usepackage{dcolumn} 
\usepackage{bm}

\begin{document}

\preprint{APS/123-QED}

\title{Miniaturized on-chip spectrometer enabled by electrochromic modulation}

\author{Menghan Tian\textsuperscript{1,2}}
\author{Baolei Liu\textsuperscript{1,2}}
 \email{liubaolei@buaa.edu.cn}
\author{Zelin Lu\textsuperscript{1}}
\author{Yao Wang\textsuperscript{1}}
\author{Ze Zheng\textsuperscript{1}}
\author{Jiaqi Song\textsuperscript{1}}
\author{Xiaolan Zhong\textsuperscript{1}}
 \email{zhongxl@buaa.edu.cn}
\author{Fan Wang\textsuperscript{1}}
 \email{fanwang@buaa.edu.cn}
 
\affiliation{%
 \textsuperscript{1}School of Physics, Beihang University, Beijing, 100191, China\\
\textsuperscript{2}These authors contributed equally to this work.
}%

\date{\today}

\begin{abstract}
Miniaturized on-chip spectrometers with small footprints, lightweight, and low cost are in great demand for portable optical sensing, lab-on-chip systems, and so on. Such miniaturized spectrometers are usually based on engineered spectral response units and then reconstruct unknown spectra with algorithms. However, due to the limited footprints of computational on-chip spectrometers, the recovered spectral resolution is limited by the number of integrated spectral response units/filters. Thus, it is challenging to improve the spectral resolution without increasing the number of used filters. Here we present a computational on-chip spectrometer using electrochromic filters that can be electrochemically modulated to increase the efficient sampling number for higher spectral resolution. These filters are directly integrated on top of the photodetector pixels, and the spectral modulation of the filters results from redox reactions during the dual injection of ions and electrons into the electrochromic material. We experimentally demonstrate that the spectral resolution of the proposed spectrometer can be effectively improved as the number of applied voltages increases. The average difference of the peak wavelengths between the reconstructed and the reference spectra decreases from 14.48 nm to 2.57 nm. We also demonstrate the proposed spectrometer can be worked with only four or two filter units, assisted by electrochromic modulation. This strategy suggests a new way to enhance the performance of miniaturized spectrometers with tunable spectral filters for high resolution, low-cost, and portable spectral sensing, and would also inspire the exploration of other stimulus responses such as photochromic and force-chromic, etc, on computational spectrometers.
\end{abstract}

\maketitle


\section{Introduction} 

Spectrometers have been widely used in fundamental scientific research, industrial inspection, and consumer electronics\cite{1,2,3}. Traditional benchtop spectrometers offer high spectral resolutions by using bulky light-dispersive components and long optical path lengths. As a counterpart, miniaturized spectrometers, which have reduced footprint, weight, and cost, are in high demand for portable optical sensing or lab-on-a-chip systems\cite{4,5,6,7}. Recently, on-chip spectrometers utilizing reconstruction algorithms have emerged as a prominent research focus. This trend is attributed to their streamlined optical hardware by leveraging the power of computational techniques\cite{7,8,9,10,11,12,13,14,15,16,17,18,19,20,21,22,23,24,25,26,27,28,29,30}. Such spectrometers typically employ random spectral filters or spectral response units, instead of bulky dispersive optics (e.g. grating and prism)\cite{31}, narrowband filters\cite{32}, or microelectromechanical systems (MEMS)\cite{33,34}. Various approaches have been employed to achieve computational spectrometers,  including quantum dot filter arrays\cite{13,14,15}, photonic crystal arrays\cite{16,17}, compositionally engineered nanowires\cite{18,29}, and in situ perovskite modulation\cite{19}.

Computational on-chip spectrometers use reconstruction algorithms to evaluate the target spectrum by considering the pre-calibration characteristics of used filters and their response to the tested light. More spectral response units result in better recovered spectral resolution. In other words, improving the number of integrated filters would enhance the performance of computational spectrometers, or the resolution is limited by the number of integrated filters. To develop a computational spectrometer without increasing its size, we turn to harness the power of electrochromism. Electrochromism is a phenomenon that reversibly changes the color or transmission spectrum, and other optical properties of materials as a response to the drive of an external electrical voltage. The basis for such a change comes from the reversible redox reactions that occur when cations and electrons are embedded in, or detached from the electrochromic materials\cite{35,36,37,38}. Electrochromic devices (ECD) have electrically tunable spectral response characteristics with high color change sensitivity over a wide spectral range, suggesting that the sampling numbers in computational spectrometers can possibly be improved by applying electrical stimulation on spectral filters, without increasing the number of filters or spectrometers’ size.

In this work, we report an electrochromic computational on-chip spectrometer (ECOS) that can utilize voltage-induced spectral modulation to increase the sampling numbers for enhanced performance of recovered spectra. The electrochromic filter array, which has a footprint of about $1cm \times 1cm$, is integrated on the top of a complementary metal–oxide–semiconductor (CMOS) sensor. The spectral response functions of the filters are tuned by applying different voltages. We investigated the reconstructed spectrum performance of monochromatic and complex spectra by ECOS at different voltages. The ECOS can resolve the spectrum that has two peaks separated by 10nm with improved spectral resolution as the increase of applied voltages. We also demonstrate that the on-chip spectrometers with only two or four filters are possible with the aid of electrochromics. Our ECOS concept suggests a novel strategy to improve the sampling numbers in miniaturized spectrometers, for better spectral performance, in a cost-effective way. It can be further extended by applying the spectral modulations in the other stimulus responses such as photochromic and force-chromic, etc.

\section{Results}

\begin{figure*}
\includegraphics[width=\textwidth]{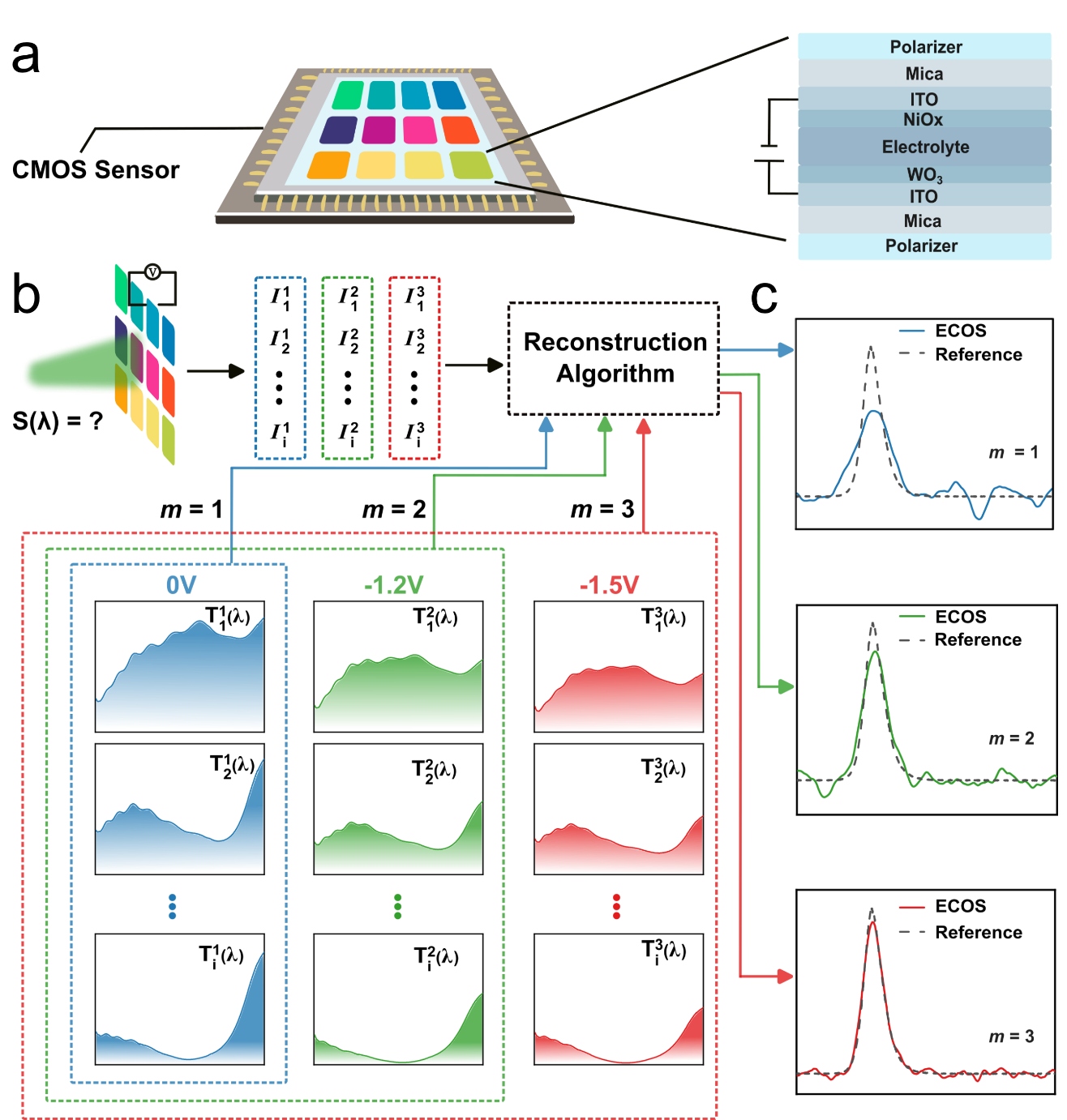}%
\caption{Scheme and principle of the proposed electrochromic computational on-chip spectrometer (ECOS). (a) Schematic of ECOS, which consists of an electrochromic filter array on top of the CMOS sensor. The electrochromic filter array consists of an ECD embedded in two polarizers with different polarization angles. The ECD consists of five layers of ITO/$\mathrm{WO_{3}}$/electrolyte/$\mathrm{NiO_{x}}$/ITO on a mica substrate. (b) Principle of ECOS. The transmission spectra of different electrochromic filters at different voltages (‘0V’, ‘-1.2V’, and ‘-1.5V’). S($\lambda$) refers to the unknown light for testing. (c) The reconstructed spectra at different voltages. The \textit{m} denotes different detection modes: ‘\textit{m}=1 (blue)’ refers to the mode of using only ‘0V’, ‘\textit{m}=2 (green)’ refers to the mode of using ‘0V’ and ‘-1.2V’, and ‘\textit{m}=3 (red)’ refers to the mode of using ‘0V’, ‘-1.2V’  and ‘-1.5V’.}
\label{fig:1}
\end{figure*}

The electrochromic filter array consists of two polarizers with different polarization angles and an ECD embedded between them.  The electrochromic filter array, which has a size of $1cm \times 1cm$, is integrated directly on top of the COMS sensor to form the ECOS, as shown in Fig. 1a.  Each electrochromic filter has a unique transmission spectrum due to the chromatic polarization effect induced by the birefringent property of ECD, modulated by the polarization angles of the two polarizers. In addition, the transmission spectrum of each electrochromic filter can also be tuned by applying different voltages to the ECD. The ECD is the thin-film stacks combined with five layers of ITO/$\mathrm{WO_{3}}$/electrolyte/$\mathrm{NiO_{x}}$/ITO deposited on mica substrate with birefringent properties, in which $\mathrm{WO_{3}}$ and $\mathrm{NiO_{x}}$ were chosen as the cathode electrochromic layer and the anode electrochromic layer, respectively, due to their excellent electrochromic properties\cite{35,36,37,38}. More details of the electrochromic principles can be found in Supplementary Information.

The conventional on-chip spectrometers rely on engineered spectral filters or chromatic detection units that have fixed spectral responsivity. Increasing the number of filters \textit{n} in a given spectral range improves the spectral resolution, however, the number of filters is limited for a computational spectrometer with a certain size in practical applications. The electrochromic filters have different spectral responses at different voltages, which means that the sampling number can be increased by tuning the transmission spectra. Each filter with a different polarization angle ($P_1$) has a unique transmission spectrum, and the same filter with different voltages also exhibits different spectral characteristics and constitutes the response function (Fig. 1b, red dot line box). This pre-calibrated response function together with the measured unknown transmission intensity at different voltages are processed by algorithms to reconstruct the unknown spectrum.

The operation principle of the proposed ECOS is shown in Fig. 1(b). $T_{i}^{k}(\lambda)$ is the transmission spectrum produced by the \textit{i}th electrochromic filter under the kth (\textit{k} = 1, 2, 3) applying voltages. Here, the three used voltages were 0V, -1.2V, and -1.5V, respectively. The set of transmitted light intensity $I_{i}^{k}$, corresponding to the response of each filter, is measured by the CMOS sensor simultaneously:

\begin{equation}
    I_{i}^{k}=\int_{\lambda_{min}}^{\lambda_{max}}S(\lambda)T_{i}^{k}(\lambda)\eta(\lambda)d\lambda
\end{equation}
where \textit{i} = 1, 2,… \textit{n}, \textit{n} is the number of filters, $\lambda$ is the wavelength, S($\lambda$) is an arbitrary incident light spectrum, $\eta(\lambda)$ is the quantum efficiency of the sensor,  $\lambda_{min}$ and $\lambda_{max}$ are the minimum and maximum detectable wavelength, respectively. The \textit{m} (\textit{m} = 1, 2, 3) in Fig. 1c denotes different detection modes, where \textit{m} = 1, 2, 3 represent the usage of ‘0 V’, ‘0 V, -1.2 V’ and ‘0 V, -1.2 V, -1.5 V’ sampling voltages, respectively. The measured intensity $I_{i}^{k}$ at \textit{m} voltage groups, together with the pre-calibrated response function $T_{i}^{k}(\lambda)$ of \textit{m} voltage groups are used to reconstruct the spectrum $S(\lambda)$ (Fig. 1 b-c).

\begin{figure*}
\includegraphics[width=\textwidth]{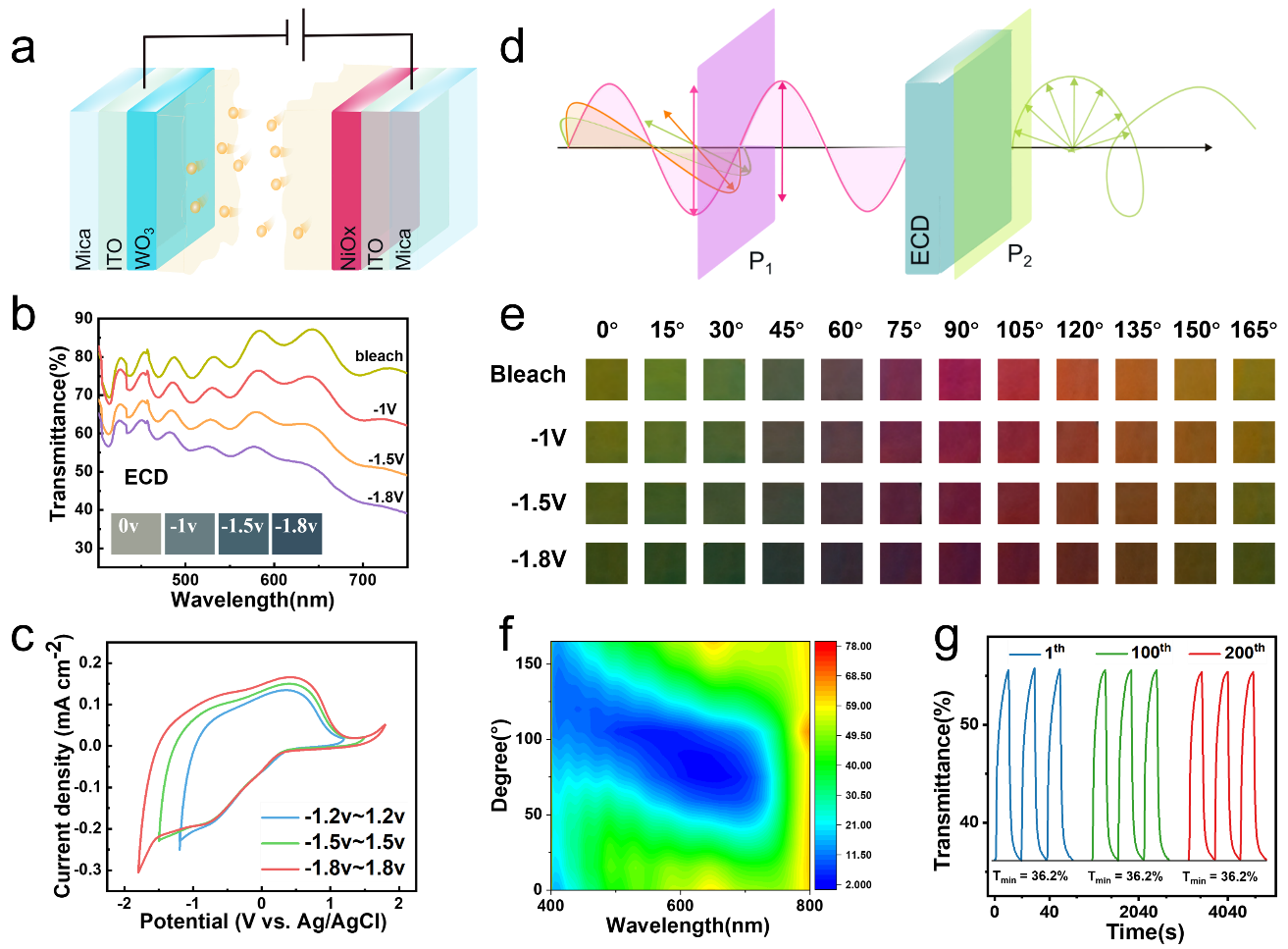}%
\caption{Optical characteristics of the ECD. (a) Schematic diagram of the ECD structure.Transmittance spectra (b) and cyclic voltammograms (c) with the different potential ranges for ECD at a sweep rate of 100 mV/s. (d) Schematic of the polarization response of ECD spectra. (e) Color variation of ECD corresponds to different polarization angles and voltages. (f) Transmittance spectra of ECD (bleached state) at different polarization angles. (g) Transmittance evolution curves of ECD at the wavelength of 550 nm.}
\label{fig:2}
\end{figure*}

\begin{figure*}
\includegraphics[width=\textwidth]{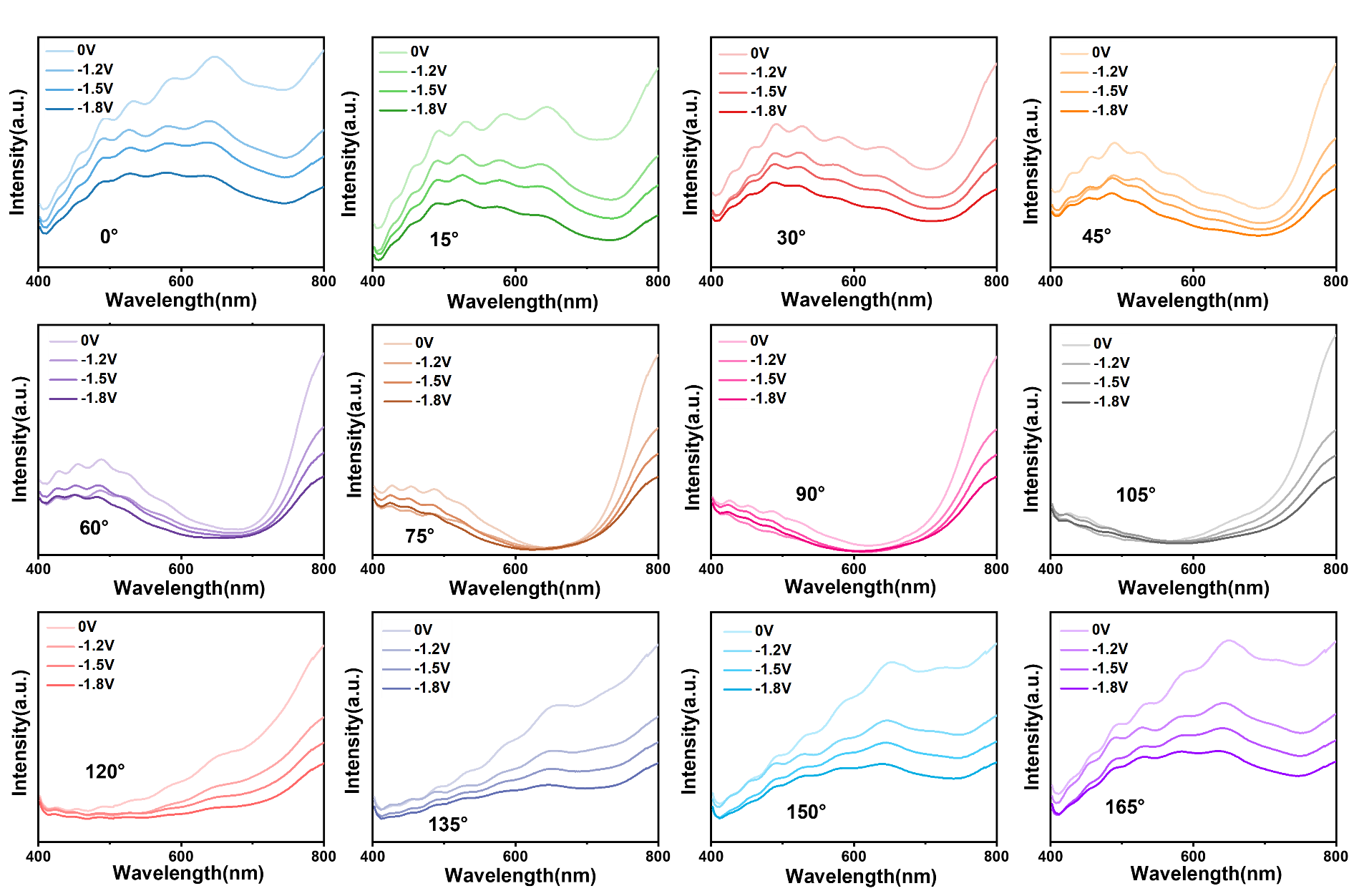}%
\caption{Measured transmission spectra of the ECD at different voltages for different polarization angles.}
\label{fig:3}
\end{figure*}

\begin{figure*}
\includegraphics[width=17.5cm]{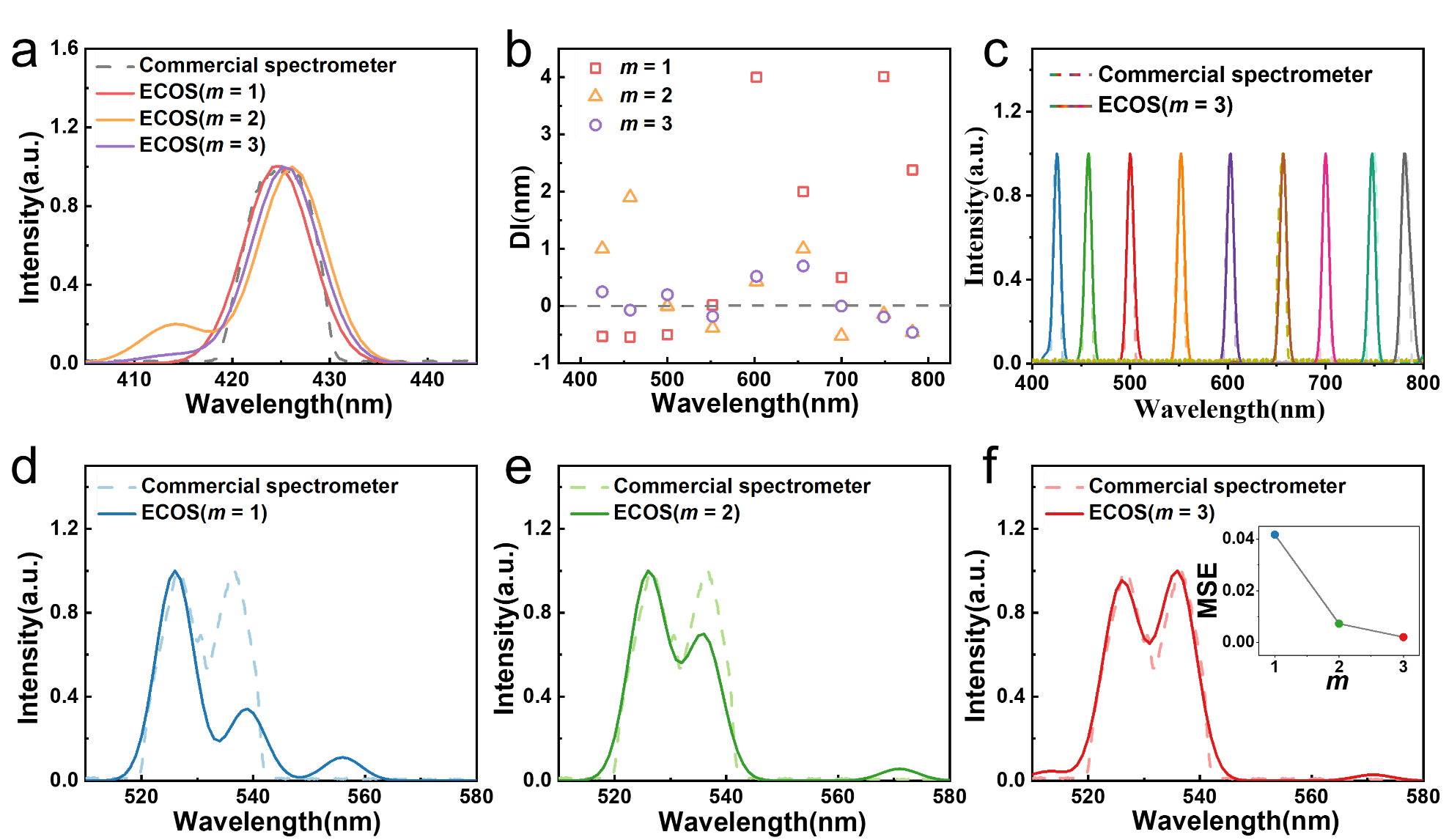}%
\caption{Tunability of the spectra resolution under different detection modes. (a) Reconstruction of a single spectral peak at 425 nm. (b) The differences in the peak wavelength between reconstruction and reference spectra. (c) Reconstructed spectra of nine monochromatic lights over the entire operating wavelength range reconstructed with \textit{m} = 3. (d-f) Reconstructions of the spectrum with two mixed peaks. The inset of (f) shows the mean square error (MSE) of different modes of \textit{m}. Here \textit{m} denotes different detection modes: ‘\textit{m} = 1 (blue)’ refers to the usage of only ‘0V’, ‘\textit{m} = 2 (green)’ refers to the usage of ‘0 V’ and ‘-1.2 V’, and ‘\textit{m} = 3 (red)’ refers to the usage of ‘0 V’, ‘-1.2 V’  and ‘-1.5 V’.}
\label{fig:4}
\end{figure*}

Figure 2a is the schematic diagram of ECD with the five-layer structure. Figure 2c elucidates the cyclic voltammetry (CV) curves of the device at different voltage ranges at a scan rate of 100 mV/s. All CV curves show similar shapes and none of them showed polarization. The device can operate normally and stably in the large voltage range from -1.8 V to 1.8 V. It is also possible to increase the operating voltage as required. The envelope area increases with the expansion of the voltage window, which means an increase in charge. The larger charge allows more ions to participate in the color change process, resulting in lower transmission spectra (Fig. 2b). The color change is shown in the inset, and it can be clearly seen that the color changes gradually from transparent to blue as the charging voltage increases. The ECDs are positioned between two polarizers (Fig. 2d). Due to the chromatic polarization effect, refer to the birefringence and the dispersion of ECD, different polarization angles of the first polarizer $P_1$ can provide different transmission colors. Figure 2e illustrates the color change of the ECD at various polarization angles. Unlike traditional ECDs that change color from transparent to indigo blue, combined with the electrochromic effect, the device exhibits a variety of colors such as green, purple, and red as the polarization angle changes from 0° to 180°. The colors at the same angle change gradually as the voltage changes. The corresponding transmittance spectra without applied voltage are shown in Fig. 2f, the transmittance spectra at different angles are highly variable and can be used for spectral modulation in computational spectrometers. Figure S4 provides evidence that the device has a broad variation of CIE coordinates in different angles with a wide range of color spans and rich color modulations. Notably, the device has eminent optical stability, since the transmittance intensity remains unchanged for more than 200 cycles as shown in the in-situ transmittance evolution curve at 550 nm of ECD (Fig. 2g).

The performance of the computational spectrometer depends on the variability of its wavelength-dependent optical response. To further observe the spectral response of the electrochromic filter, we measure the transmission spectra of the ECD with different polarization angles at different voltages, as shown in Fig. 3. With a wavelength range of 400 - 800 nm and a step of 1 nm, these spectra vary with different applied voltages of ‘0 V, -1.2 V, -1.5 V’ and an additional ‘-1.8 V’. In this work, we choose the first three voltages for the verification of the proposed spectrometer. However, other voltages such as the ‘-1.8 V’ may be further explored for increased spectral accuracy of reconstructions, as the corresponding spectra still change.

To investigate the spectral resolution of ECOS, we measured the spectrum of monochromatic lights (Fig. 4). Figure 4a shows the reconstructed spectra of monochromatic light at 425 nm with working mode (\textit{m} = 1, 2, and 3). The three reconstructed spectra show good agreement with the input spectrum (measured by commercial spectrometers), with more sampling data providing better reconstruction results (e.g. mode \textit{m} = 3). Figure 4b shows the quantified difference ($\delta\lambda$)  between the reconstructed peak wavelength and reference peak wavelength. As the working mode \textit{m} varies from 1 to 3, the sampling number increases (12, 24, 36 for \textit{m} = 1, 2, 3, respectively), and the resultant average difference $\delta\lambda$ is reduced from 14.48 nm (\textit{m} = 1) to 5.83 nm (\textit{m} = 2), 2.57 nm (\textit{m} = 3). We further present the reconstructions of nine representative narrowbands’ transmission spectra in the range of 400-800 nm at the mode of \textit{m} = 3 in Fig. 4c, which agree with the reference spectra measured from the commercial spectrometer (USB4000-UV-VIS-ES, Ocean Optics). Furthermore, we measured a complex spectrum that has two mixed peaks separated by 10 nm at around 530 nm. The reconstruction of \textit{m} = 3 (Fig. 4f) shows high-quality agreement with the reference spectrum, while the ones of \textit{m} = 1, 2 show the loss of part of the spectrum (Fig. 4e, f). We quantify the performance of ECOS by the mean square error (MSE, lower is better), which is defined as follows:

\begin{equation}
MSE = \frac{1}{M}\sum_{i=1}^{M}(\hat{P}[\lambda_{i}]-P[\lambda_{i}])^{2} , i = 1,2,…,M
\end{equation}
where \textit{M} is the spectral length, and $\hat{P}[\lambda_{i}]$ and $P[\lambda_{i}]$ are the normalized reference spectrum and normalized reconstructed spectrum, respectively. The MSE values of the reconstructed spectra from the three working modes of \textit{m} = 1, 2, and 3 are 0.04177, 0.00725, and 0.00207 respectively (in the inset of Fig. 4f). This indicates that the proposed ECOS has reduced artifact and improved spectral resolution with the usage of additional voltage. More example reconstructions of the complex spectrum that has two mixed peaks can be found in Fig. S5.

To further demonstrate the ability of our spectrometer, we examined the reconstructions with reduced filter numbers of only four or two, as shown in Fig. 5. Firstly, we used four filter units (0°, 30°, 60°, 135°) to reconstruct three representative monochromatic lights, which have peaks of 450 nm, 550 nm, and 652 nm (Figs. 5a-c). We further quantify them by using MSE in Fig. 5e. For the three monochromatic lights, the MSE value of the reconstructed spectra gradually decreases as \textit{m} increases. For example, the MSE value of the reconstructed spectra of 550 nm monochromatic light is reduced from 0.14213 (\textit{m} = 1) to 0.04321(\textit{m} = 2), 0.01738 (\textit{m} = 3). Similarly, we also used our spectrometer to measure the spectra of green LED light by only using these four filter units (\textit{n} = 4). The reconstructed spectra of the green LED by the three working modes are in great agreement with the reference spectrum (Fig. 5d), with all the MSE values below 0.02 (Fig. 5e). Their good agreement observed may be attributed to the broader spectral features of the LED light that can be retrieved by using the four filter units without applying the voltage. The MSE value of \textit{m} = 2 (0.01868) is slightly higher than that of \textit{m} = 1 (0.01247), due to measurement or calibration noise.

We next demonstrated the possibility of reconstructing the spectrum of the green LED light by using only two filter units (60° and 135°). Due to the lower resolution provided by two units, mode 1 and mode 2 cannot reconstruct the spectrum with the correct peak position, with the MSEs of 0.0562 and 0.05251, respectively (Fig. 5f). Encouragingly, the reconstructed spectrum by \textit{m} = 3 is good agreement with the reference spectrum with MSE of 0.00607, which suggests the potential feasibility of computational spectrometer using few filter units or only a single unit.  Overall, the proposed ECOS spectrometer has demonstrated a strategy to improve the spectral resolution by applying the voltage to the ECD, especially for ultracompact microspectrometers that have limited filter numbers. If a priori information about the spectral characteristics of the target spectrum is available, the desired spectral resolution can be obtained by selecting the appropriate number of voltages. In other words, if a lower spectral resolution is enough for practical applications, we may choose to not use voltage modulation (\textit{m} = 1) for simple procedures. If a higher spectral resolution is desired, we can choose more voltage modulation (e.g. \textit{m} = 2 or 3, or even higher).

\begin{figure*}
\includegraphics[width=\textwidth]{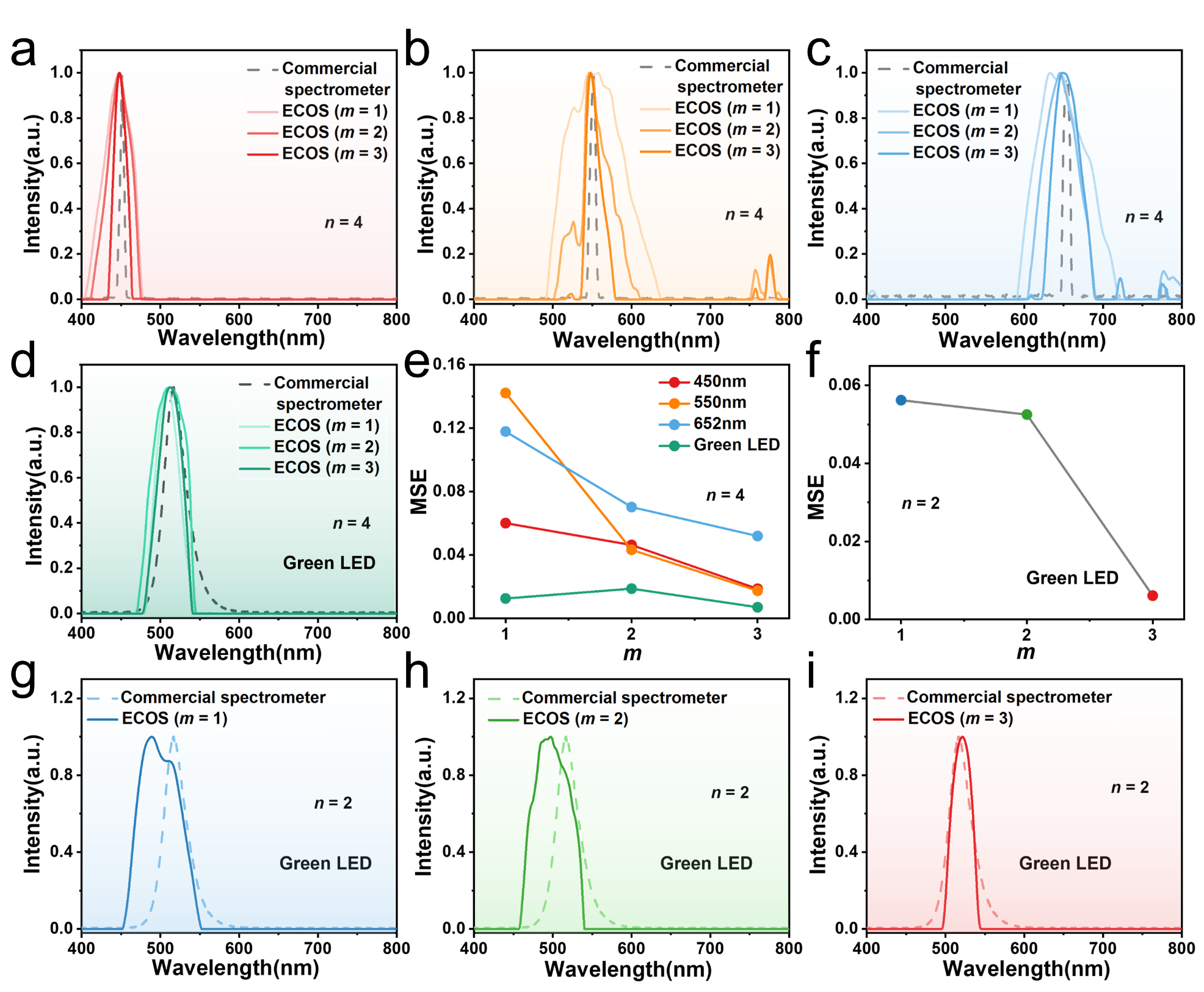}
\caption{Spectral resolution of reconstructed spectra with reduced numbers of filter units \textit{n}. (a-c) The reconstructed spectra of three representative monochromatic light (peaks of 450 nm, 550 nm, and 652 nm) by using only four filter units (0°, 30°, 60°, 135°). (d) The reconstructed spectra of the green LED light by using four filter units (0°, 30°, 60°, 135°). (e) The MSE values of reconstructed spectra in (a)-(d). (f) The MSE values of reconstructed spectra of green LED light under three modes of voltage utilization by only using two filter units. (g-i) The reconstructed spectra of green LED light by using only two filter units (60°, 135°). Here \textit{m} denotes different detection modes: ‘\textit{m} = 1 (blue)’ refers to the usage of only ‘0V’, ‘\textit{m} = 2 (green)’ refers to the usage of ‘0V’ and ‘-1.2V’, and ‘\textit{m} = 3 (red)’ refers to the usage of ‘0V’, ‘-1.2V’  and ‘-1.5V’.}
\label{fig:5}
\end{figure*}

\section{Discussion}

In this work, we have demonstrated an on-chip spectrometer (ECOS) that uses an electrochromic  filter array to effectively improve the sampling number and reconstruction quality in the range of 400 nm-800 nm, since the transmission spectra of the filters can be tuned by applying additional voltages. The reconstructed spectra of multiple narrowband and broadband signals have shown improved quality by measuring their MSE values, compared with the results from traditional methods without voltage modulation. Moreover, we also demonstrated it is possible to obtain incident spectra by only using four or two filters for developing computational spectrometers with fewer filter units or one single unit (few-pixel or single-pixel spectrometers).

A further study can be conducted to achieve hyperspectral imaging with the proposed ECOS spectrometer, such as capturing the 3D spatio-spectral cubes by using the spatial point scanning method\cite{18} or compressive single-pixel imaging method\cite{39,40,41,42,43}. Optimization algorithms based on deep learning can be used to improve the accuracy and speed of the reconstruction\cite{44}. By optimizing electrochromic materials and device types, not only the operating bandwidth of the spectrometer can be broadened, but also the switching speed can be improved. For example, the preparation of polyaniline (PANI) on plasmonic nanoslit arrays or the  PEDOT-MeOH device with double-gyroid-structured can increase the electrochromic coloring speed to 9 ms or 23 ms, respectively\cite{45,46,47}. Other filter-based on-chip spectrometers can also be electrically modulated, such as directly integrating the electrochromic devices with quantum dots filters\cite{13} or photonic crystal arrays\cite{16,17}. This work also paves the way for using other modulation methods, such as photo modulation\cite{48}, force modulation\cite{49}, and thermal modulation\cite{50,51} to develop novel on-chip spectrometers. We, therefore, anticipate that the proposed ECOS concept will open new opportunities for developing cost-effective, easy-to-prepare on-chip spectrometers or integrated hyperspectral imaging systems.

\section{Materials and Methods}

\subsection{Fabrication of electrochromic device}
The ITO, $\mathrm{WO_{3}}$ films, and the ITO, $\mathrm{NiO_{x}}$ films were prepared stepwise on mica substrates by pulsed direct current (DC) reactive magnetron sputtering. The targets used in the experiment are a tungsten metal target, nickel metal target, and ITO target with a purity of 99.99\% and a diameter of 10 cm. The mica substrates were first ultrasonically cleaned with acetone, anhydrous ethanol, and deionized water for 15 minutes, respectively, and then dried in an oven at 60°C. The $\mathrm{WO_{3}}$ films were prepared with the following parameters: argon to oxygen ratio was 3:1, the sputtering pressure was 2 Pa, the sputtering time was 30 min, and the sputtering power of 300W. The $\mathrm{NiO_{x}}$ films were prepared with the following parameters: argon to oxygen ratio was 94:6, the sputtering pressure was 3 Pa, the sputtering time was 30 min, and the sputtering power was 150W. The ITO films were prepared with the following parameters: argon-to-oxygen ratio was 78.4:1.6, the sputtering pressure was 0.3 Pa, the sputtering time was 20 min, and the sputtering power was 180W. All samples were prepared at room temperature. The $\mathrm{NiO_{x}}$ film was used as the anodic electrochromic layer, $\mathrm{WO_{3}}$ films were used as the cathodic electrochromic layer to form an electrochromic device, and the electrolyte was formed by mixing 1 mol/L PC/$\mathrm{LiClO_{4}}$ solution and UV-glue in a volume ratio of 2:1. The device was placed under UV light for 15 minutes. The device structure can be expressed as Mica/ITO/$\mathrm{WO_{3}}$/polymer electrolyte/$\mathrm{NiO_{x}}$/ITO/Mica after simple encapsulation. The voltammetry cycle (CV) was measured with an electrochemical workstation (CHI660E, Shanghai Chenhua) and the transmittance of the sample was measured by an ultraviolet-visible spectrometer (TU-1810, PERSEE).

\subsection{Fabrication of electrochromic filter}
A polarizer of $1cm \times 1cm$ is placed on the back side of the electrochromic device of $1cm \times 1cm$, and 12 polarizers of $0.2cm \times 0.3cm$ are placed on the front side of the device. The polarization angles of the front polarizer and the back polarizer are 0°, 15°,…,165°, with an interval of 15°. The electrochromic filter was directly integrated in front of the COMS camera (acA1920-150um, Basler).

\subsection{Fabrication of electrochromic filter}
The xenon lamp (GLORIA-X150A, Zolix) is used to provide continuous spectra, and monochromatic light output controlled by a monochromator (Omni-$\mathrm{\lambda}$300, Zolix) with a step of 1 nm, with a range of 400 nm to 800 nm. The spectrum of the light source is measured by the commercial spectrometer (USB4000-UV-VIS-ES, Ocean Optics) for reference purposes. The unknown light was first coupled into an optical fiber and then the light illuminated the proposed spectrometer after collimation. An \emph{f} = 50 mm lens is used for light collimation. The exposure time of the camera was 83 ms or 200 ms for the narrowband light or the green LED light, respectively.

\subsection{Computational reconstruction}
\subsubsection{Calibration}
An electrochemical workstation provides the voltage to drive the electrochromic device. In the calibration process, the voltage of the driver device was chosen as 0 V, -1.2 V, and -1.5 V. The measured intensities of each filter unit from read out of the COMS camera (acA1920-150um, Basler) form the spectral responsiveness $T_{i}^{k}(\lambda)$. All measurements were performed at room temperature.
\vspace{\baselineskip}
\subsubsection{Reconstruction algorithms}
The reconstruction algorithm of Fig. 4 is the constrained least square solution with the adaptive Tikhonov regularization method by minimizing the residual norm with a regularization factor for its excellent performance on narrowband light\cite{7,18}. The reconstruction algorithm of Fig. 5 is the total variation minimization by augmented lagrangian and alternating direction algorithm (TVAL3)\cite{52}.

\bigskip

\begin{acknowledgments}
This work was supported by the National Natural Science Foundation of China (U23A20481, 62075004, 11804018, 62275010); Beijing Municipal Natural Science Foundation (1232027, 4212051); China Postdoctoral Science Foundation (2022M720347); the International Postdoctoral Exchange Fellowship Program (YJ20220241); the Fundamental Research Funds for the Central Universities.
\end{acknowledgments}

\section*{Author contributions}
M.T., B.L., X. Z., and F. W. conceived the idea. M.T. and B.L. performed the measurements and conducted the spectrum reconstruction. M.T. fabricated the samples. M.T. and B.L. analyzed the results and prepared the manuscript. X. Z. and F. W. supervised the project. All the authors participated in the discussion and confirmed the final manuscript.

\section*{Conflict of interest}
The authors declare no competing interests.

\section*{Supplementary information}
See Supplementary Information for supporting content.

\nocite{*}

\bibliography{ref}

\end{document}